# Optimizing time-spectral solution of initial-value problems

J. Scheffel and K. Lindvall

*Department of Fusion Plasma Physics, School of Electrical Engineering,
KTH Royal Institute of Technology, SE-100 44 Stockholm, Sweden
E-mail: jan.scheffel@ee.kth.se*

## Abstract

Time-spectral solution of ordinary and partial differential equations is often regarded as an inefficient approach. The associated extension of the time domain, as compared to finite difference methods, is believed to result in uncomfortably many numerical operations and high memory requirements. It is shown in this work that performance is substantially enhanced by the introduction of algorithms for temporal and spatial subdomains in combination with sparse matrix methods. The accuracy and efficiency of the recently developed time spectral, generalized weighted residual method (GWRM) is compared to that of the explicit Lax-Wendroff method and the implicit Crank-Nicolson method. Three initial-value PDEs are employed as model problems; the 1D Burger equation, a forced 1D wave equation and a coupled system of 14 linearized ideal magnetohydrodynamic (MHD) equations. It is found that the GWRM is more efficient than the time-stepping methods at high accuracies. For time-averaged solution of the two-time-scales, forced wave equation GWRM performance exceeds the finite difference methods by an order of magnitude both in terms of CPU time and memory requirement. Favourable scaling of CPU time and memory usage with the number of temporal and spatial subdomains is demonstrated for the MHD equations.

# 1. Introduction

In time-spectral methods for time-dependent ordinary and partial differential equations, a spectral representation is employed for the temporal domain. As an alternative to standard finite differencing, this approach has been studied by a number of authors [1-24]. It is sometimes held, however, that computing the solution simultaneously over all space-time is inefficient [1]. In this work, we will show that high efficiency can indeed be obtained through the use of optimizing methods, including spatial and temporal subdomains.

The focus is here on the recently developed Generalized Weighted Residual Method (GWRM), where truncated Chebyshev expansions are employed [25,26]. Similarly as for other time-spectral approaches, the CFL condition and other grid causality conditions associated with time marching algorithms are eliminated. Although the problems to be solved typically are causal, the method is acausal in the sense that the time dependence is calculated by a global minimization procedure (the weighted residual formalism) acting on the time integrated problem. Recall that, in standard WRM, initial value problems are transformed into a set of of coupled linear or nonlinear ordinary differential equations for the time-dependent expansion coefficients [27]. These are solved using finite differencing techniques.

In the GWRM, not only temporal and spatial but also physical parameter domains may be treated spectrally using Chebyshev polynomials, being of interest for carrying out parameter scaling dependence in a single computation. How this works becomes clear as the method is briefly described in the next section.

Returning to the question of efficiency, most of the GWRM computational effort is spent in solving the system of linear or nonlinear (depending on the type of problem solved) algebraic equations for the Chebyshev series coefficients. Iterative root solvers require either the computation of an inverse matrix or the solution of an equivalent matrix equation. As a simple example consider solution of a 1D initial-value partial differential equation, employing Chebyshev polynomials of order $K$ and $L$ in time and space, respectively. Then $\Omega = [(K+1)(L+1)]^3$ numerical operations are typically required for matrix inversion and $\Omega/3$ operations using $LU$ decomposition for solving the corresponding matrix equation [28]. Should large modal numbers $K$ and $L$ be necessary for sufficient resolution of the computational domain, the corresponding large number of operations may indeed prohibit any positive comparison with finite difference methods. Furthermore the memory requirements can be shown to scale as $[(K+1)(L+1)]^2$. It is clear that measures need be taken to reduce these numbers.

The paper is arranged as follows. In the next section, a short introduction to the GWRM is provided. In section 3 several methods for improved GWRM efficiency will be presented. These will, in turn, be implemented as we compare the efficiency of the GWRM versus explicit and implicit methods in section 4. The paper ends with discussion and conclusion.

## 2. The generalized weighted residual method (GWRM)

We may write a system of parabolic or hyperbolic initial-value partial differential equations symbolically as

$$\frac{\partial \boldsymbol{u}}{\partial t} = \boldsymbol{D}\boldsymbol{u} + f \tag{1}$$

where $\boldsymbol{u} = \boldsymbol{u}(t, \boldsymbol{x}; \boldsymbol{p})$ is the solution vector, $\boldsymbol{D}$ is a linear or nonlinear matrix operator and $f = f(t, \boldsymbol{x}; \boldsymbol{p})$ is an explicitly given source (or forcing) term. Note that $\boldsymbol{D}$ may depend on *both* physical variables ($t$, $\boldsymbol{x}$ and $\boldsymbol{u}$) and physical parameters (denoted $\boldsymbol{p}$) and that $f$ is assumed arbitrary but non-dependent on $\boldsymbol{u}$. Initial $\boldsymbol{u}(t_0,\boldsymbol{x};\boldsymbol{p})$ as well as (Dirichlet, Neumann or Robin) boundary $\boldsymbol{u}(t,\boldsymbol{x}_B;\boldsymbol{p})$ conditions are assumed known.

Our aim is to determine a spectral solution of Eq.(1), using Chebyshev polynomials [29] in all dimensions. For simplicity, we restrict the discussion to a single equation with one spatial dimension $x$ and one physical parameter $p$. Thus the solution is approximated by (a prime denotes that zeroth order terms are multiplied by ½)

$$u(t, x; p) = \sum_{k=0}^{K}{}' \sum_{l=0}^{L}{}' \sum_{m=0}^{M}{}' a_{klm} T_k(t) T_l(x) T_m(p) \tag{2}$$

The Chebyshev polynomials of the first kind (henceforth simply referred to as Chebyshev polynomials) are defined by $T_n(x) = \cos(n \arccos(x))$. These are real ordinary polynomials of degree *n*, orthogonal in the interval [-1,1] over a weight $w_x = (1 - x^2)^{-1/2}$. Thus $T_0(x) = 1$, $T_1(x) = x$, $T_2(x) = 2x^2 - 1$ and so forth.

As in standard WRM, a residual *R* is defined as

$$R \equiv u(t, x; p) - [u(t_0, x; p) + \int_{t_0}^{t} \{Du + f\} dt'] \tag{3}$$

The coefficients $a_{klm}$ of the Chebyshev series are subsequently determined from the set of algebraic equations being generated by *R* from the requirement that the residual should satisfy the Galerkin WRM defined over the full computational domain

$$\int_{t_0}^{t_1} \int_{x_0}^{x_1} \int_{p_0}^{p_1} R T_q(t) T_r(x) T_s(p) w_t w_x w_p \, dt \, dx \, dp = 0 \tag{4}$$

To this end, the right hand terms of Eq. (1) have all been expanded in Chebyshev polynomials. The resulting algebraic equations are solved using the iterative solver SIR [30], which features improved convergence characteristics as compared to Newton's method with linesearch. Details of the GWRM procedure, including handling of boundary conditions, can be found in [25,26].

All computations are performed using the computer mathematics programme Maple. The GWRM is easily coded in languages like Matlab or Fortran, but absolute computational speed is not important for the comparisons with finite difference methods made here; rather it is important that all comparisons are carried out within the same computational environment.

## 3. Improving efficiency

An early implementation of the GWRM was compared with finite difference methods for solving two elementary initial-value problems in [25]. Studies of accuracy and efficiency were made for the nonlinear 1D Burger equation and a linear, forced 1D wave equation, respectively.

The *1D Burger equation*, being related to problems in fluid mechanics and magnetohydrodynamics (MHD), is

$$\frac{\partial u}{\partial t} = -u\frac{\partial u}{\partial x} + v\frac{\partial^2 u}{\partial x^2} \tag{5}$$

where $v$ can be interpreted as (kinematic) viscosity. For comparisons, we use an exact solution of this equation [25]. It was found in [25] that, for specified accuracy, the Burger equation was solved about two times faster for $v = 0.01$ by the Lax-Wendroff method than by the GWRM and about four times faster with a semi-implicit method, advancing the linear diffusive term with the Crank-Nicolson scheme and the nonlinear convective term explicitly.

The *1D forced wave equation* being solved is

$$\frac{\partial^2 u}{\partial t^2} = v\frac{\partial^2 u}{\partial x^2} + f(t,x) \tag{6}$$

$$u(t,0) = u(t,1) = 0$$
$$u(0,x) = \sin(n\pi x)$$
$$\frac{\partial u}{\partial t}(0,x) = \alpha A sin(\beta x)$$

where the forcing function is $f(t,x) = A(v\beta^2 - \alpha^2)\sin(\alpha t)$. This equation has the exact solution $u(t,x) = \cos(n\pi v^{0.5}t)\sin(n\pi x) + Asin(\alpha t)\sin(\beta x)$, featuring two time scales with the driving term time scale much longer than the intrinsic time scale; the respective ratio is $R = \alpha/(n\pi\sqrt{v})$. The primary aim was here to average out the fast time scale behaviour in order to generate approximate solutions following the slower time scale. For similar accuracy, the GWRM was here about 10 times faster than Lax-Wendroff and 30 times faster than Crank-Nicolson.

In the following, we will present algorithm improvements that substantially enhance the performance of the GWRM for these two problems. Furthermore, GWRM performance improvements for a third, advanced problem will be studied; the set of 14 (7 complex), linearized ideal MHD equations modelling the stability of a magnetically confined plasma.

How then, is the GWRM made more efficient? The measures that can be taken fall essentially into two categories: a) optimal adaption of SIR to the GWRM and b) streamlining of the GWRM itself. Below we present the ideas and algorithms that have been developed for these categories; performance results will be given in the next section.

**3.1 SIR optimization**

ODE's and PDE's can be solved globally by the GWRM scheme given in section 2 using single spatial and temporal domains. High resolution then requires high modal numbers $K$ and $L$ (we let $M = 0$ in this paper) which in turn results in a large set of $N = (K+1)(L+1)$ nonlinear or linear algebraic equations to be solved simultaneously by SIR. A natural step to avoid the corresponding cubic and quadratic dependencies on $N$ for the number of operations and memory storage, respectively, would be to divide the physical domain into coupled subdomains in space and time.

Substantial CPU time would be saved if the subdomain equations could be computed independently to some extent. Attempts to update the spatial domains independently at each iteration, using previous iterates for boundary conditions only, was however found to be only partially successful [31]. Convergence requires for this approach that the initial iterates are chosen very close to the solution. In fact it has been shown both theoretically and computationally that iteration convergence, in terms of a limited maximum norm, usually requires a formulation that, by some procedure, couples all equations in each iteration [30]. In the following this latter, 'dependent' subdomain approach is thus employed.

The root solver SIR [30] is at the core of the GWRM. We will now discuss what measures have been taken to optimize SIR for GWRM use.

**S1. Matrix and vector numerical package.** It is important that the computational environment includes efficient packages for standard operations on vectors and matrices. In Maple, the transition from the linalg to the LinearAlgebra package resulted in faster handling of the matrix equations. Certain packages, like VectorCalculus, should not be called globally since they slow down computations.

**S2. Solution of matrix equations.** In SIR, the matrix equation $x = A(x - \varphi) + \varphi$ is solved iteratively, where the vector **x** contains the Chebyshev coefficients of the solution $u$, $\varphi$ is a vector with components that are functions of the coefficients, and $A$ is a linear matrix operator being computed to provide optimal convergence at each iteration. To determine $A$, a linear matrix equation involving the system Jacobian $J \equiv \partial(x - \varphi)/\partial x$ need be solved. A large fraction of the GWRM CPU time lies here. Using LU decomposition solution of this system, instead of inversion of *J,* a dependence $\Omega/3$ rather than $\Omega$ for the number of operations is obtained for large matrices. For small matrices, however, inversion turns out to be faster, thus there is an option to chose either method.

**S3. Choice of equation solver mode.** For many problems, SIR can be run as Newton's method since sufficient convergence is achieved and fewer iterations are needed. For improved convergence, SIR default settings [30] are preferably used.

**S4. Effect of A matrix on convergence.** When solving *linear* algebraic equations, ***A*** need be computed only for the first domain, provided that the domains are equidistant in time, and can then be re-used for the following time domains. This fact is extremely useful when dividing the temporal domain of the problem into subdomains. *Nonlinear* PDEs usually require at least 5-10 iterations. For the last few iterations, however, the ***A*** matrix is nearly constant. Thus substantial CPU time is saved by re-computing ***A*** in the first few iterations only; beautiful houses can be built with ugly scaffolds also.

**S5. Band matrix methods.** Sparse, band-shaped Jacobian matrices *J* occur in problems where many spatial subdomains are employed because only neighbouring domains are analytically coupled. Maple has built-in algorithms that automatically handle sparse matrix equations efficiently.

**S6. *J* matrix differentiation.** The Jacobian *J* is obtained exactly by analytical differentiation of $\varphi$. This is a tedious procedure that, without optimization, may require more than 50 % of the total GWRM CPU time for matrices of dimension about 3000 or higher. By implementing algorithms that differentiate the non-zero band matrix elements only, favourable scaling with the number of spatial subdomains is obtained for very large matrices.

**S7. Spatial and temporal subdomain influence on $\varphi$.** In particular for nonlinear problems, the components of $\varphi$ may be lengthy and complex, thus being time-consuming to differentiate analytically. Significant speed is gained by the use of spatial and temporal subdomains, since then the same global accuracy may be obtained using lower order Chebyshev polynomial expansions in each subdomain, resulting in more manageable $\varphi$ vectors for differentiation.

**S8. Choice of initial vector $x = x_0$.** As for all iterative methods, SIR convergence strongly depends on the choice of initial vector $x_0$. The closer to the solution, the faster the convergence. In GWRM computations, $x_0$ is typically taken to be the initial condition or, when multiple time domains are used, the solution for the end of the previous time interval. Thus, if the temporal length is reduced, the solution vector $x$ will arbitrarily approach the initial guess $x = x_0$. Hence, GWRM convergence is always guaranteed. In some computations particularly well conditioned choices of can be made. For example in numerical weather prediction, several scenarios are computed with slightly different initial conditions in order to provide ensemble results. Rapid GWRM convergence can then be reached by using solutions $x$ from previous computations as $x_0$ [32].

## 3.2 GWRM optimization

Next follows a discussion on the measures taken to optimize the GWRM.

**G1. Spatial and temporal subdomains.** The use of spatial and temporal subdomains implies that the same accuracy can be retained with lower order Chebyshev polynomials. Optimistically, if this order could be reduced to half by halving the interval, a speed gain of about a factor 4 would be obtained because of the cubic dependence on the number of modes and that two, rather than one, intervals need to be computed. In reality, the story is more complicated and there is an optimum subinterval length [32]. For the time domain this means time intervals that may be a factor of 100 longer than the time steps of, for example, Runge-Kutta methods and for the spatial domain the optimum Chebyshev order is typically much higher than those of finite element methods. As mentioned regarding SIR optimization, a large number of spatial subdomains is favourable for efficiency since corresponding Jacobian will become a very sparse band matrix due to that only immediately neighbouring domains will contribute to non-zero near-diagonal matrix elements.

**G2. Overlapping spatial subdomains.** It is preferable to use overlapping spatial subdomains in Chebyshev spectral methods as compared to matching function and functional derivative values at borders. Standard is two-point overlap ("handshake"). The reason is that the Chebyshev spectral space representation of derivatives is sensitive to the values of higher order coefficients, which values are quite approximative both during early iterations and for solutions that do not need be precise. The amount of overlap can be chosen arbitrarily; very small values (order $10^{-6}$ of the spatial domain) are usually favourable for high accuracy. The number of overlap points required to preserve boundary condition information across the spatial domain is a function of the number of first order PDE's that are solved [31].

**G3. Adaptive temporal subdomains.** Time overlap is only used for the temporal domains when it enhances convergence, since accuracy generally is negatively affected. Adaptive time interval length, however, greatly enhances efficiency. Best results have been obtained by starting with a relatively long time interval; if convergence is not reached, the time interval is reduced and a new computation is performed. The algorithm regularly strives to increase the time interval length, which procedure is very forceful in smooth computational terrain. It may be mentioned that this algorithm is very robust since Chebyshev polynomials are limited to values in the interval [-1,1]; thus higher order Chebyshev coefficients directly measure convergence.

**G4. Time parallellization.** The use of spatial subdomains opens up the possibility for performing strongly parallel computations in each time interval. In an approach termed 'the common boundary condition method' (CBC) we solve the physical equations of each subdomain in parallel for each iteration, whereas the global computation only involves the boundary equations that connect the domains. This promising procedure is relatively complex and will be reported elsewhere.

**G5. Clenshaws algorithm.** Nearly all GWRM computations take place in spectral space. The computation of a Chebyshev series however, which may be needed for example when handling overlapping temporal domains, is inaccurate at higher modal numbers. Clenshaw's algorithm [28] allows accurate high order representations and should be used instead.

**G6. End conditions.** Since the GWRM is an acausal algorithm, initial conditions can be traded for end conditions for possible improvement of numerical stability. This potential avenue is, so far, only explored for some simple cases with neutral result.

## 4. Results

Early implementations of the GWRM have been compared with finite difference methods with respect to convergence, accuracy and efficiency for the two model problems discussed above [25]. Efficiency enhancement of the GWRM, employing the ideas of section 3, will now be demonstrated for these cases including an advanced problem related to fusion plasma confinement.

**4.1 Accuracy – the Burger equation**

In [25] Burger's equation was solved by the GWRM for $v = 0.01$ with the parameters $T = 10$, $N_t = 1$, $K = 9$, $N_s = 2$, $L = 7$ using an algorithm where the spatial subdomains were solved independently at each iteration, and coupled thereafter. Run parameters were CPU time 2.48 s and memory use 182 MB. This algorithm is often numerically unstable [31] and is therefore not reported elsewhere in this paper. The unoptimized code in [25], with the spatial subdomains simultaneously ("dependently") solved at each iteration now required 5 iterations for an accuracy of $1.0 \cdot 10^{-3}$, using 14.1 s CPU time and 192 Mb. The new, optimized code is substantially more efficient, using 1.27 s and 37.1 MB.

For comparison, the same accuracy is obtained with the second order Lax-Wendroff method [28] in 2.37 s, using 234 MB of memory. The spatial grid needs 70 points for accuracy whereas 1000 time steps are needed to satisfy the dominating CFL criterion $\Delta t \leq (\Delta x)^2/(2v)$ for this problem [25]. A semi-implicit method, advancing the linear diffusive term using the Crank-Nicolson scheme and the nonlinear convective term explicitly was also implemented. Again employing $\Delta x = 1/70$, only 500 time steps, 0.47 CPU s and 37.1 MB of memory were required for an accuracy of $1.0 \cdot 10^{-3}$.

In summary, for an accuracy of $1.0 \cdot 10^{-3}$ the optimized GWRM solution of the Burger equation for $v = 0.01$ required about half the CPU time of the explicit Lax-Wendroff method and only 15 % of the memory. The semi-implicit method needed the same amount of memory as the GWRM but was about two times faster. In this section accuracy is studied, so we now turn to comparisons for higher accuracy.

Using the optimized GWRM, again for $v = 0.01$, an increased accuracy of $1.0 \cdot 10^{-4}$ was obtainable for $T = 10$, $N_t = 5$, $K = 6$, $N_s = 5$, $L = 7$, requiring 4 iterations for each time interval, 6.72 CPU s and 88.3 MB. The Lax-Wendroff method needed 57.4 CPU s and 1430 MB, using $\Delta x = 1/200$ and 8100 time steps. Corresponding parameters for the semi-implicit method was 28.6 CPU s, 456 MB, $\Delta x = 1/400$ and 4500 time steps. Increasing accuracy to $1.0 \cdot 10^{-5}$, the GWRM provides a solution for $T = 10$, $N_t = 12$, $K = 6$, $N_s = 8$, $L = 7$, with 3 iterations for each time interval, in 32.3 CPU s using 195 MB of memory. This accuracy could neither be achieved with the Lax-Wendroff nor the Crank-Nicolson method within 180 CPU s or below 3000 MB of memory. As an example, $2.0 \cdot 10^{-5}$ accuracy was found for the latter method using $\Delta x = 1/900$ and 22000 time steps in 472 CPU s for 3390 MB memory use.

Thus it is concluded that for $v = 0.01$ and an accuracy of $1.0 \cdot 10^{-4}$ the optimized GWRM solution required 12 % of the CPU time and 6 % of the memory of the Lax-Wendroff method. When compared to the Crank-Nicolson method, the numbers become 23 % and 19 % for CPU and memory requirements, respectively. The GWRM consequently strongly outperforms both finite difference methods for higher accuracies. For lower accuracies the finite difference methods become more competitive.

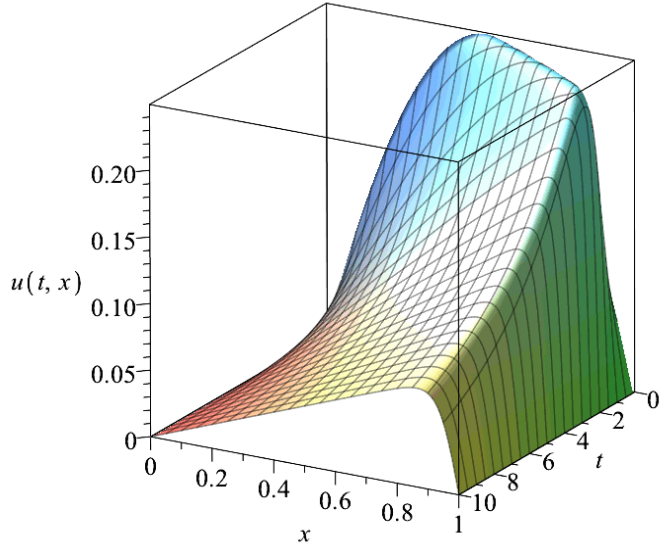

**Figure 1. GWRM solution of Burger's equation; $v = 0.003$. For parameters see text.**

It is well known that spectral methods often are less efficient for problems where shocks or steep gradients need be resolved. This is confirmed for the stiffer 1D Burger case $v = 0.003$. A steep gradient towards $x = 1$ develops due to convection, as can be seen in Figure 1. The GWRM provides a $7.0 \cdot 10^{-4}$ accurate solution for $T = 10$, $N_t = 5$, $K = 6$, $N_s = 9$, $L = 7$, with maximum 4 iterations for each time interval, in 17.4 CPU s using 181 MB of memory. The Lax-Wendroff method requires, for the same accuracy, 2.75 CPU s and 180 MB, with $\Delta x = 1/80$ and 1000 time steps. Corresponding parameters for the semi-implicit method are 4.62 CPU s and 187 MB, using $\Delta x = 1/80$ and 4000 time steps. For an accuracy of $1.0 \cdot 10^{-4}$ the GWRM needs $T = 10$, $N_t = 10$, $K = 6$, $N_s = 20$, $L = 7$, with maximum 4 iterations for each time interval, in 153 CPU s using 306 MB of memory. The Lax-Wendroff method uses 73.2 CPU s and 1420 MB for the parameters $\Delta x = 1/200$ and 10000 time steps, whereas the semi-implicit method uses 106 CPU s and 2040 MB for the parameters $\Delta x = 1/300$ and 20000 time steps. Thus it is again seen that for high accuracy the GWRM becomes more efficient, primarily with regards to memory consumption.

Of particular interest is GWRM CPU time and memory scaling with $N_t$ and $N_s$. Using the case mentioned at the beginning of this section we have performed scans where $N_t \in [1,15]$ and $N_s \in [1,15]$. It was found that CPU time scales as $N_t^{1.0} N_s^{1.43}$ and memory usage as $N_t^{0.0} N_s^{1.08}$ (for $N_t > 2$). These scalings represent a substantial improvement as compared to the cubic and quadratic scalings with $N_t N_s$ for CPU time and memory, respectively, that hold for unoptimized code without subdomains (assuming $KN_t$ and $LN_s$ global modes would be used instead).

Finally, it is worthwhile to consider which of the measures S1-S8, G1-G6 that contributes most to the improved GWRM performance. Clearly the simultaneous use of temporal and spatial subdomains (G1,G2) is important through the avoidance of high numbers of global temporal and spatial modes. The CPU time linear dependence (and memory independence) on $N_t$ is expected, whereas band matrix methods (S5), and also measures S1-S4, S6-S7, contribute to the weak dependence on $N_s$. The present problem is easily solved by SIR, which converges also in Newton mode, being quite insensitive to the choice of initial vector $\mathbf{x}_0$ (S8). Measures G3-G6 were unimportant here. We may mention, however, that measure G3, automatic time interval adaption, may improve efficiency substantially in certain problems; for example in a solution of three coupled, time-dependent and chaotic ODE's it leads to GWRM efficiency beyond that of fourth order Runge-Kutta methods [32].

### 4.2 Efficiency – a forced wave equation

The forced 1D wave equation studied in [25] features two distinct time scales; a slow time scale associated with the driving function and a fast system time scale. A major reason for developing the GWRM was its potential to average out small scale oscillations, thus enhancing efficiency by using a reduced number of spectral modes to follow slow time scales only. Explicit methods are here hampered by the limiting CFL conditions associated with signals travelling at the fast time scale. Indeed it was found in [25] that, for the problem studied, the GWRM is about 3 times faster than the Lax-Wendroff method and 30 times faster than Crank-Nicolson's method. The latter method is slowed down by the need to solve matrix equations at each time step since multiple equations are solved.

Focusing on efficiency in finding smoothed, time-averaged solutions, accuracy is here determined by comparison with the slow time scale part of the exact solution, that is the second term of $u(t,x) = \cos(n\pi v^{0.5}t)\sin(n\pi x) + A\sin(\alpha t)\sin(\beta x)$.

The optimized GWRM code solves the case of [25] (with $T = 30$, $N_t = 1$, $K = 6$, $N_s = 1$, $L = 8$ for $A = 10$, $\alpha = 2\pi/T$, $\beta = 3\pi$, $n = 3$) to an accuracy of 0.08 in 0.212 CPU s using 36.1 MB of memory. The Lax-Wendroff method solution of [25] (with $\Delta x = 1/30$ and 900 time steps) uses 0.828 s and 69.1 MB for an accuracy of 0.30.

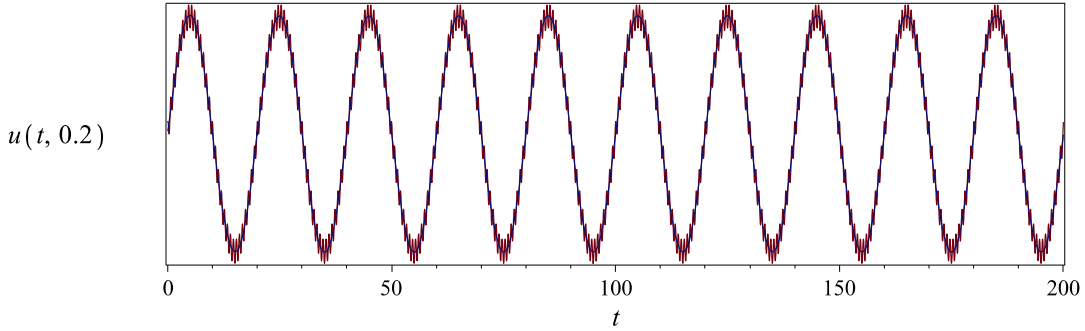

**Figure 2. GWRM time-averaged solution versus time *t* of the wave equation at *x* = 0.2, compared to exact, oscillatory solution. For parameters, see text.**

Being a hyperbolic equation, the wave equation is not well suited for the use of implicit methods because of the problem of resolving phase at time steps larger than that given by the CFL condition. Here, however, the emphasis is rather on time-averaged accuracy and efficiency, thus it is of interest to see how an implicit method like Crank-Nicolson's performs. This method has now been optimized in relation to [25]. For the case $\Delta x = 1/30$ using 100 time steps, a limited time-averaged accuracy of 0.87 was achieved employing 1.16 s and 64.1 MB. The Lax-Wendroff method is thus preferable of the two finite difference methods in this case, in spite of being explicit. The GWRM, however, is more accurate and much faster than both the finite difference methods.

The case above features a single wavelength of the slow time scale. In practical situations many period, slow time scale solutions often are of interest. In Figure 2 we show a GWRM solution of the same problem above for 10 periods (with $T = 200$, $N_t = 10$, $K = 6$, $N_s = 2$, $L = 8$ for $A = 10$, $\alpha = 20\pi/T$, $\beta = 3\pi$, $n = 3$). A global accuracy of 0.22 was obtained using 2.66 CPU s and 83.2 MB of memory. Using $N_s = 1$ (a single spatial domain) nearly the same accuracy was obtained in 1.08 s, using 66.7 MB.

Comparing with the finite difference methods, Lax-Wendroff obtains the same accuracy with $\Delta x = 1/50$ and 10000 time steps (CFL limit) using as much as 15.8 CPU s and 442 MB. The Crank-Nicolson method features low accuracy because of phase drift and is thus outperformed by the Lax-Wendroff method.

## 4.3 GWRM solution of the linearized, perturbed ideal MHD equations

Magnetohydrodynamic stability is a necessary condition for magnetically confined fusion plasmas. Theoretically, the stability of a specified plasma equilibrium may be tested by linearizing the ideal MHD equations

$$\frac{\partial \rho}{\partial t} + \nabla \cdot (\rho \boldsymbol{u}) = 0 \qquad (7)$$

$$\rho \frac{d\boldsymbol{u}}{dt} = \boldsymbol{j} \times \boldsymbol{B} - \nabla p$$

$$\boldsymbol{E} + \boldsymbol{u} \times \boldsymbol{B} = \boldsymbol{0}$$

$$\frac{d}{dt}(p\rho^{-\Gamma}) = 0$$

$$\nabla \times \boldsymbol{E} = -\frac{\partial \boldsymbol{B}}{\partial t}$$

$$\nabla \times \boldsymbol{B} = \mu_0 \boldsymbol{j}$$

Having specified the boundary conditions (in this case in circular cylindrical geometry), a perturbation is applied and the time dynamics is investigated for possible exponential growth, in which case the equilibrium is unstable. Details are given in [25], where it is shown that 14 coupled scalar (7 complex) PDEs need be solved simultaneously. Notable is that the evolution, in the unstable case, will be given by the competition of a number of unstable modes with different number of radial nodes. As the fastest growing mode (with zero radial nodes) starts to dominate, memory of the initial perturbation is gone.

The stability of two equilibria will be studied here, applying the GWRM. The first is that of [25]; A) a screw-pinch equilibrium with radially constant current density profile and axial magnetic field $B_{0z} = 0.05$ (normalized units, erroneously given as 0.2 in [25]); the second case B) is a pure z-pinch so that $B_{0z} = 0$. The azimuthal perturbation has Fourier mode number $m = 1$ and axial mode number given by $k = 10$. Both equilibria are strongly unstable to this perturbation, featuring exponential growth rates of order unity (normalized to the Alfvén time). A difficulty for the GWRM is thus to polynomially resolve the exponential growth in time. In order for the dominant mode to develop, the equations typically need be solved for times $T = 10$ or more. For benchmarking, GWRM results are compared with an eigenvalue shooting code [33]. All computations are run in Maple on the same platform.

First we note that the CPU time and memory requirement for case A, earlier discussed in [25], is 26.0 s and 444 MB respectively. For this case 5 time intervals were used for a single spatial domain; furthermore temporal $K$ and spatial $K$ maximum mode numbers were both 5. Employing the improvements of sections 3.1 and 3.2, the CPU time is reduced to 4.44 s and memory to 89.8 MB. Both cases gave the same result; growth rates = 0.83 within 1 % error and eigenfunctions within approximately 2 % error.

Of particular interest is dependence on number of time intervals $N_t$ and number of spatial domains $N_s$. Since a linear equation is solved, the $\boldsymbol{A}$ matrix need be solved only for the first time interval (see S4). For the case above, the first time interval needs 1.49 s for full solution, whereas succeeding time intervals on average require only 0.68 s, thus a 54 % reduction. The CPU time scaling with $N_t$ for these time intervals is linear. Memory requirements are essentially independent of $N_t$.

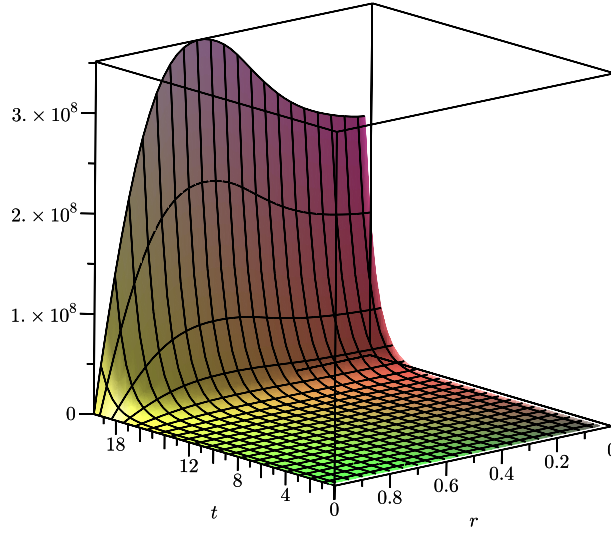

**Figure 3. GWRM solution showing exponential time evolution (a.u.) of perturbed radial velocity $u_{1r}$ versus time $t$ and radial coordinate $r$, for unstable equilibrium B).**

For case B) the parameters $T = 20$, $N_t = 3$, $K = 5$, $N_s = 1$, $L = 5$ were used for a run that took 11.1 s, using 105 MB memory, with 15 % maximum error in eigenfunction $u_r$. Using $N_s = 2$ (with $1.0 \cdot 10^{-6}$ overlap), the CPU time increased to 26.2 s and memory to 336 MB, whereas eigenfunction error decreased to 5 %; see Figure 3. The correct growth rate 1.04 was achieved within 1 % error. Increasing the number of spatial domains, the CPU time scaling $N_s^{1.49}$ was obtained, in stark contrast to the unoptimized scaling $N_s^3$. The memory scaling was found to be $N_s^{1.69}$ (rather than $N_s^2$) as a result of memory use unrelated to SIR.

## 5. Discussion

The ambition of this work has been to evaluate the performance of optimized implementations of the time-spectral method GWRM as compared to finite difference methods in time. In early work [25] some example PDEs were solved. It was found that the explicit Lax-Wendroff and implicit Crank-Nicolson methods were both somewhat more efficient in finding accurate solutions to the 1D Burger equation, whereas the GWRM outperformed the finite difference methods in tracing the longer time scale behaviour of a PDE representing a forced wave equation. An advanced problem in MHD, including 14 simultaneous PDE's, was accurately solved but it was realized that the cubic CPU time and quadratic memory dependence on the total number of Chebyshev modes limits the performance of the method. Subdomains in time and space need be used for advanced problems. In [31] it was found that the spatial domains could be decoupled during the iterations for some problems, which dramatically increases performance, but the method is not universally stable and usually requires good initial guesses (by, for example, using short time intervals) for the root solver SIR. Thus results from fully coupled spatial subdomains are reported here. As described in section 3, a number of measures to enhance efficiency both for the GWRM itself, but also for SIR, have been developed. Returning to the earlier model problems, using the new algorithms, we can now report strongly enhanced performance. Of primary importance are the improved CPU time and memory scalings, where the usage of sparse matrix methods play central roles.

Let us now estimate the requirements for solution of an advanced 2D problem. Primarily, GWRM efficiency depends on the solution of a matrix equation in SIR for determining the matrix **A**. Without employing subdomains in time and space, the dimension $N$ of this matrix is determined by the number of simultaneous equations to be solved $N_e$, and the number of Chebyshev modes $(K, L_x, L_y)$; thus $N = N_e K L_x L_y$. With $N_e = 5, K = 100, L_x = 50, L_y = 50$, we obtain $N = 1.3 \cdot 10^6$. Standard Gauss elimination requires $\Psi = O(N^3)$ operations for each SIR iteration. Thus, for this case, $\Psi \approx 2 \cdot 10^{18}$ operations, which would call for high performance computers.

A substantial improvement in efficiency comes from the introduction of subdomains in time and space. We let $L_x = N_x L_{sx}$, $L_y = N_y L_{sy}$ and $K = N_t K_s$. Here $N_x$ and $N_y$ denote the number of spatial subdomains in the *x*- and *y*-directions, respectively and $N_t$ is the number of temporal subdomains. $L_{sx}, L_{sy}$, and $K_s$ denote the number of Chebyshev modes used for each domain. In the unoptimized case we have approximately $N = N_e K_s N_x N_y L_{sx} L_{sy}$ and $\Psi = O(N_t N^3)$. Letting $N_t = 10$, $N_e = 5$, $K_s = 10$, $N_x = 10$, $N_y = 10$, $L_{sx} = 5$, $L_{sy} = 5$ we find $N = 1.3 \cdot 10^5$ and $\Psi = 2.0 \cdot 10^{16}$, Clearly, spatial optimization is of the essence. The scalings found from the optimizations presented in this paper substantially improves the situation. Using the optimized dependency $O((N_x N_y)^{1.45})$ obtained in this work rather than $O((N_x N_y)^3)$ we find $\Psi = 1.6 \cdot 10^{13}$, a substantial reduction. A gigahertz table top computer could thus solve the problem within a few hours.

The scalings above are indeed validated for the 1D problems considered in this paper; taking into account the $N_e$ dependence good agreement is obtained with the CPU times used.

Turning to 3D problems, we may assume a further scaling of the number of operations with $N_z^{1.45} L_{sz}^3$. Thus for a problem with $N_z = 10$ and $L_{sz} = 5$, using the above parameters, we have $N = 6.5 \cdot 10^6$ and $\Psi = 5.6 \cdot 10^{16}$, which is not prohibitive for high performance computers.

GWRM efficiency can, however, be further enhanced. In recent work, to be published elsewhere, the number of simultaneous global spatial equations to be solved by SIR is reduced to the boundary equations (external plus internal) only. The physics equations of each spatial subdomain are solved locally at each iteration, and strong time parallelization is possible. The resulting improved scalings are particularly important for problems with multiple spatial dimensions.

In this paper we have not employed automatic adaption of the time intervals (G4). This method has been proven to be very efficient when the GWRM was used for solving a set of chaotic differential equations in time, typical for numerical weather prediction [32]. Time adaption lead to accurate GWRM solution of this problem at least as efficient as fourth order Runge-Kutta methods. Thus automatic time interval adaption, global solution of boundary equations only and parallelization will be interesting further paths of development of the GWRM for applications on advanced problems.

## 6. Conclusions

The time-spectral generalized weighted residual method (GWRM) replaces the time differencing of standard methods for solving differential equations with a Chebyshev polynomial representation in time. Unoptimized use of the method is hampered in efficiency by the cubic dependence of the number of operations on the total number of modes. Several measures for enhancing efficiency, primarily sparse matrix methods, have been studied when employing multiple temporal and spatial domains.

It was found that Burger's 1D equation, with viscosity parameter $v = 0.01$, was solved significantly faster and more accurate by the GWRM than by the explicit Lax-Wendroff and the implicit Crank-Nicolson finite difference methods for accuracies of order $1.0 \cdot 10^{-4}$ or higher. For lower values of viscosity, where a steep gradient develops, the finite difference methods perform somewhat better than the GWRM. Furthermore, it was found that GWRM CPU time scales as $N_t^{1.0} N_s^{1.43}$ and memory usage as $N_t^{0.0} N_s^{1.08}$, where $N_t$ and $N_s$ are the number of time intervals and spatial subdomains, respectively. This is a significant improvement of the $N_s^3$ and $N_s^2$ scalings, respectively, in the unoptimized case.

The slower time scale of a forced wave equation, solved by all three methods, is found and followed by the GWRM much faster and using less memory than the finite difference methods.

For the ideal MHD stability problem solved it was found that the measures S1-S8, G1-G5 of section 3 yielded a more than five-fold increase in efficiency. Being a linear problem, for which information from the first time interval can be reused, the CPU time for further time intervals becomes halved. A CPU time scaling with spatial subdomains $N_s^{1.49}$ was obtained; a substantial reduction of the unoptimized scaling $N_s^3$. The memory scaling was somewhat improved to $N_s^{1.69}$ (as compared to $N_s^2$). The scalings enable solution of advanced 2D and 3D problems using the GWRM. A scheme that further enhances efficiency by reducing the global set of GWRM equations simultaneously solved to the external and internal boundary condition equations alone is presently developed.

In closing, it may be mentioned that all obtained GWRM solutions are analytical piece-wise polynomial expressions in time and space, thus tractable for analysis. By using Chebyshev expansions also in parameter space, scaling behaviour can be determined in a single GWRM run.